\documentclass[10pt,aps,prl,reprint,superscriptaddress,floatfix,showpacs]{revtex4-1}
\usepackage{xcolor}
\usepackage{cancel}

\usepackage{amsmath}

\usepackage{amssymb}
\usepackage{empheq}
\usepackage{braket}
\usepackage{lipsum}
\usepackage{hyperref}

\usepackage{bm}
\usepackage{graphicx}
\usepackage{hyperref}
\usepackage{natbib}
\usepackage{siunitx}
\usepackage{float}
\usepackage{soul} 
\hypersetup{
    colorlinks,
    linkcolor={blue!50!black},
    citecolor={blue!50!black},
    urlcolor={blue!80!black}
}

\usepackage{filecontents}
\begin{document}

\title[Engineering quantum interference in van der Waals heterostructures]{Engineering quantum interference in van der Waals heterostructures}
\author{Muralidhar Nalabothula}
\affiliation{Physics Department, Indian Institute of Technology Bombay, Mumbai 400076, India}
\author{Pankaj Jha}
\affiliation{Department of Applied Physics and Materials Science, California Institute of Technology Pasadena, California 91125, USA}
\author{Tony Low}
\affiliation{Department of Electrical and Computer Engineering, University of Minnesota, Minneapolis, Minnesota 55455, United States}
\author{Anshuman Kumar}
\email{anshuman.kumar@iitb.ac.in}
\affiliation{Physics Department, Indian Institute of Technology Bombay, Mumbai 400076, India}

\date{\today}
\begin{abstract}
In this paper, we present a novel route to tunable spontaneous valley coherence in heterostructures of two dimensional valleytronic materials with other layered materials hosting anisotropic polaritonic modes. We first discuss the dependence of this coherence on the conductivity tensor of the anisotropic part of a general heterostructure and on its geometrical configuration. Subsequently, we propose two implementations - one, using anisotropic plasmons in phosphorene and another with hyperbolic phonon polaritons in $\alpha-$MoO$_3$. In both these systems we show for the first time electrostatic tunability of the spontaneous valley coherence achieving unprecedented values of up to 80\% in the near to mid infrared wavelengths at room temperature. The tunability of this valley coherence shown in these heterostructures will enable the realization of active valleytronic quantum circuitry. 
\end{abstract}
\maketitle
\emph{Introduction---}
Coherent superposition of states is a fundamental feature which distinguishes quantum mechanics from its classical counterpart. As such, quantum coherence is important from a fundamental physics perspective. This property is also critical for many of the emerging applications such as quantum computation and communication. For practical applications, it is important to consider such coherence for solid state excitations. Of particular interest recently are excitons in two dimensional gapped Dirac systems such as transition metal dichalcogenides (TMDCs)\cite{RevModPhys.90.021001} and biased bilayer graphene (BLG)\cite{Ju2017}. The electronic bandstructure of such systems consists of two inequivalent yet degenerate valleys ($K$ and $K'$) in $k-$space for which the optical selection rule is sensitive to the helicity ($\sigma_{\pm}$) of the exciting photon. This selection rule is inherited by excitons associated with these two valleys when Coulomb interaction is taken into account\cite{Schaibley2016,Bussolotti2018}. There have been several proposals to use this ``valley degree of freedom'' for the development of optoelectronic devices analogous to those in the field of spintronics\cite{doi:10.1002/smll.201801483}. 

In order to harness this valley degree of freedom, it is imperative to be able to actively control the coherence between excitons in the two valleys. Such a coherence has been demonstrated in the stimulated regime utilizing an external source such as a laser\cite{Jones2013} and has been theoretically proposed in the spontaneous regime, using anisotropic metasurfaces\cite{PhysRevLett.121.116102}. The latter is of special interest since this technique allows one to generate valley coherence without the need for any external field. In this work, we theoretically demonstrate for the first time, how this spontaneous valley coherence can be achieved by creating a heterostructure of the valley material with another anisotropic polaritonic (such as plasmonic or phononic) material. Unlike previous proposals, our proposed route to spontaneous valley coherence does not require spatial patterning into nanostructures and allows for electrostatic tunability of the coherence. We begin by discussing how the spontaneous excitonic valley coherence depends on the optical conductivity of the anisotropic 2D material and its proximity to the valleytronic material. We highlight how the degree of birefringence ranging from elliptical to hyperbolic regimes influences this coherence. We derive the conditions for maximal valley coherence and discuss the influence of nonradiative losses. Following this, we expound on these ideas with two specific examples -- elliptical plasmons in phosphorene\cite{PhysRevApplied.12.014011,doi:10.1002/adom.201900996} and hyperbolic phonons in $\alpha-$MoO$_3$\cite{Ma2018}. We provide estimates of valley coherence in these heterostructures and its electrostatic tunability. Beyond these examples, an extensive library of natural birefringent 2D materials are already known\cite{doi:10.1002/inf2.12005} and our results serve as a blueprint for engineering valley coherence in 2D material heterostructures. The electrostatic tunability of the spontaneous valley coherence will open up new avenues in the development of active quantum optoelectronic devices using two-dimensional semiconductors.
\begin{figure*}[htbp]
    \centering
    \includegraphics[width=0.9\linewidth]{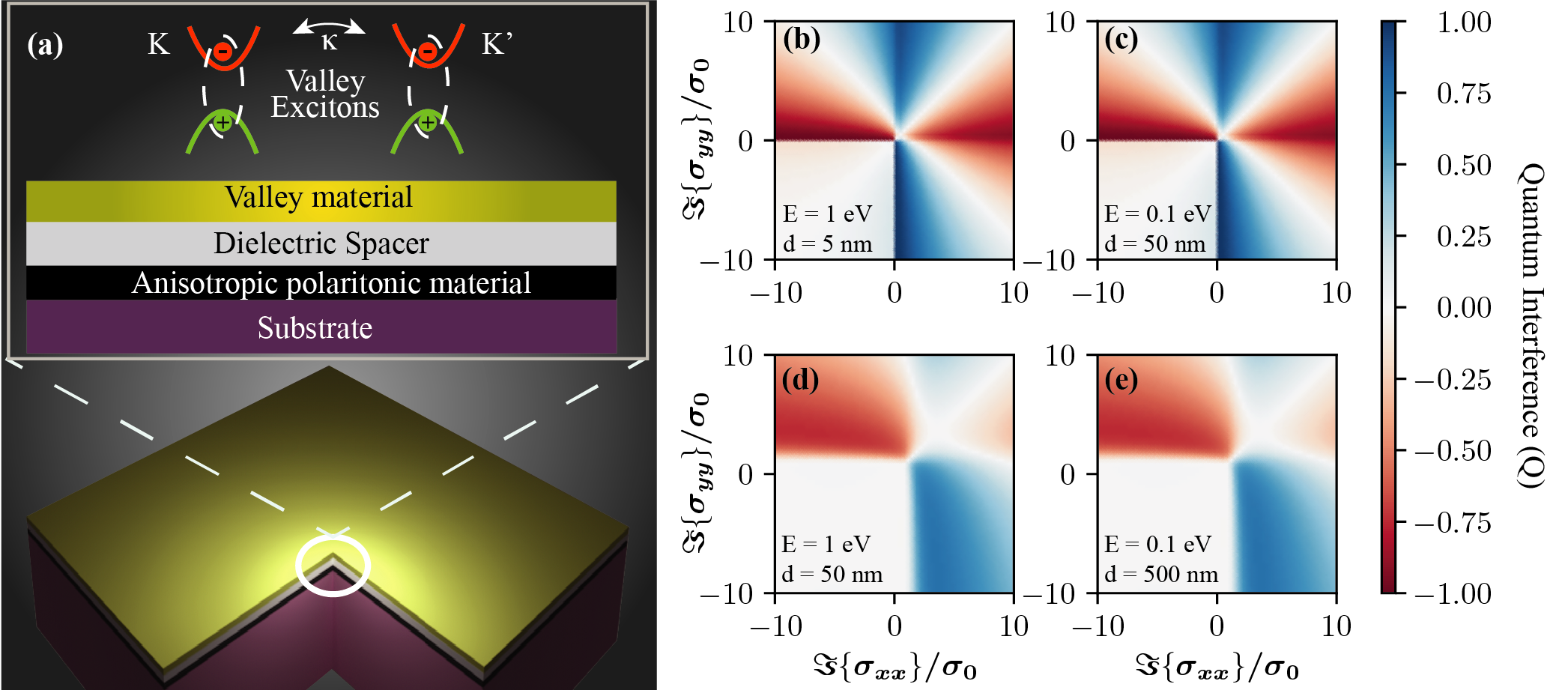}
     \caption{(a) Schematic of the heterostructure on substrate. The heterostructure is designed such that an anisotropic vacuum of electromagnetic field is created due to the 2D material, and radiation is focused back. The excitons in the valleys are modeled as in-plane circular dipoles.  (b),(d),(e) Dependence of quantum interference $(Q)$ on the diagonal conductivities for two dipole distances and exciton frequencies.} 
\label{fig:fig11}
\end{figure*}
\emph{Valley Coherence via anisotropic vacuum---}
Let us consider a valleytronic material sitting above another anisotropic polaritonic material, as shown in Fig.~\ref{fig:fig11}a. The latter can be another naturally occurring 2D material, which we will expound upon later. 
If we consider an exciton created in one valley $(K)$ and assume that the inter valley scattering rate is zero, then in free space this exciton decays (intravalley) by emitting a photon. This photon cannot excite an exciton in the orthogonal valley $(K')$ due to optical selection rule. However, in the presence of a neighbouring anisotropic layered material which induces an in-plane anisotropic vacuum near the valleytronic material, the coupling between the two valleys are allowed. Thus emission from one valley $(K)$ can excite exciton in orthogonal valley $(K')$\cite{PhysRevLett.84.5500}.

To explain this effect, we employ a three-level model comprising of two levels representing the $K$ and $K'$ valley excitons and one ground state level. Employing the formalism of Refs.\cite{1903.07426,PhysRevLett.121.116102}, we find that the spontaneous intervalley coupling rate is given by:
\begin{equation}
    \kappa_{{\nu,\Bar{\nu}}} =  \frac{6\pi\epsilon_0}{|\mu_{\nu\Bar{\nu}}|^2 k^3} \Im\{\mu^*_{\nu}\cdot \mathbf{E}_s^{\Bar{\nu}}(\mathbf{r},\mathbf{r};\omega) \}
\end{equation}
where ${\nu,\Bar{\nu}}$ denotes the two valleys \{$K, K'$\}, $\mu_{\nu}$ is dipole moment corresponding to the exciton in the valley $\nu$ and $\mathbf{E}_s^{\Bar{\nu}}(\mathbf{r},\mathbf{r};\omega)$ is scattered electric field created by the $\Bar{\nu}$ dipole and absorbed by the $\nu$ valley and $k$ is the magnitude of the wave vector. Note that the magnitude of the exciton dipole ($|\mu_{\nu\Bar{\nu}}|$) at the two valleys is equal.

Since in most valleytronic materials, the excitonic dipole at the two valleys is circularly polarized, the spontaneous coupling rate between the two valleys can be simplified as $\kappa_{\textrm{K,K'}} = \kappa = (\gamma_{x}-\gamma_y)/2$, where $\gamma_i$ is the spontaneous emission rate for an in-plane dipole sitting near the anisotropic material and oriented in the $i$th direction. It is thus straightforward to see that any photonic environment which creates an anisotropy in the spontaneous emission rate in the $x$ and $y$ directions will result in a finite valley coherence. 
In order to measure the spontaneous coherence we define Quantum Interference $Q$ as 
\begin{eqnarray}
\label{eq:eq222}
Q = \frac{\kappa}{\gamma} = \frac{\gamma_{x}-\gamma_{y}}{\gamma_{x}+\gamma_{y}}
\end{eqnarray}
where $\gamma = (\gamma_x + \gamma_y)/2$. Eq.~\ref{eq:eq222} suggests  $Q$ can take values between -1 to 1.

As a demonstration of the principle, we first explore the parameter space of arbitrary optical conductivity tensor components of the anisotropic polaritonic 2D material. To do this, we calculate the spontaneous emission rates -- $\gamma_x$ and $\gamma_y$ of an in-plane dipole, corresponding to an exciton situated at a distance $d$ above a general anisotropic 2D surface\cite{group:2870:Pur46}. 
This rate is proportional to the imaginary part of the scattered component of the dyadic Green's function (in presence of the anisotropic 2D surface), which is obtained using standard methods\cite{Lakhtakia1992,Gomez-Diaz:15}. 
Fig.~\ref{fig:fig11}(b--d) shows the dependence of Quantum Interference $Q$ on the diagonal elements of optical conductivity tensor for two distances and two exciton frequencies. Such diagrams can be constructed for other frequencies depending on the excitonic resonance for the valleytronic material in question. The four quadrants represent two hyperbolic regimes (second and fourth), that is, $\Im\{\sigma_{xx}\}\cdot \Im\{\sigma_{yy}\} < 0$ and two elliptical regimes where $\Im\{\sigma_{xx}\}\cdot \Im\{\sigma_{yy}\} > 0$\cite{PhysRevLett.114.233901}. It should be noted that the first quadrant supports elliptical plasmons whereas the third quadrant does not support any surface mode. It is clear from Fig.~\ref{fig:fig11}(b--d) that the hyperbolic regime results in the largest values of the valley coherence. This can be intuitively understood based on the fact that a hyperbolic 2D system is metallic in one direction and insulating in the other. In the lossless limit, the metallic direction can support the strongly confined TM plasmon mode whereas the insulating direction only the weakly confined TE plasmon\cite{PhysRevLett.99.016803,PhysRevB.94.195418}. The large contrast between the respective mode volumes results in a strong anisotropy in the intravalley spontaneous emission rates in the two directions, leading to valley coherence maxima reaching close to the theoretical upper limit in the hyperbolic regions. 

In the elliptical metallic region (first quadrant), we also observe the enhanced valley coherence. These regions correspond to an elliptical plasmon mode which can also produce an anisotropic spontaneous emission rate in the two directions. The contours where the valley coherence maximum occurs can again be tuned by changing the distance of the valleytronic material and the anisotropic surface. These spontaneous valley coherence maps can be used to optimize the geometry of the heterostructure and enable the selection of the appropriate optical conductivity tensor of the anisotropic 2D material to achieve the desired valley coherence. 

To interpret the features in the valley coherence map of Fig.~\ref{fig:fig11}(b--d) and find the optimal distance for the given conductivity and frequency, we employ the dispersion relation of the hetero-structure. For an anisotropic 2D material, the dispersion relation of the propagating polarition when $k_z\gg k_0$ is given by \cite{PhysRevLett.116.066804}
\begin{equation}
   \label{eq:disp}
   \sigma_{x x} \cos ^{2} \theta+\sigma_{y y} \sin ^{2} \theta=\frac{2 i \varepsilon_{0} \omega}{q}
   \end{equation}
Here $\theta = \tan^{-1}{\frac{k_y}{k_x}}$ is the propagation angle, $q = \sqrt{k_x^2+k_y^2}$ is the in-plane wave vector, $\sigma_{xx}$ and $\sigma_{yy}$ are optical conductivity tensors in $x$ and $y$ directions respectively and $\omega$ is the frequency of the dipole. Since the dipole distance $d$ should approximately match the inverse in-plane wavevector $1/q$ to couple efficiently to the corresponding polariton\cite{doi:10.1021/acs.nanolett.5b01191}, Eq.~\ref{eq:disp} becomes $\sigma_{x x} \cos ^{2} \theta+\sigma_{y y} \sin ^{2} \theta \approx 2 i \varepsilon_{0} \omega d$. 
In the lossless limit, we can assume $\sigma_{xx}  \approx i\sigma_{xx}''$ and $\sigma_{yy}  \approx i\sigma_{yy}''$ where $\sigma_{yy}''$ and $\sigma_{xx}''$ are real. The propagation angle is given by
\begin{eqnarray}
\label{eq:disp4}
   \theta = \pm\sin^{-1}\sqrt{\frac{2 \varepsilon_0 \omega d-\sigma''_{xx}}{(\sigma''_{yy}-\sigma''_{xx})}}
\end{eqnarray}
\begin{figure}
    \centering
    \includegraphics[width=0.9\linewidth]{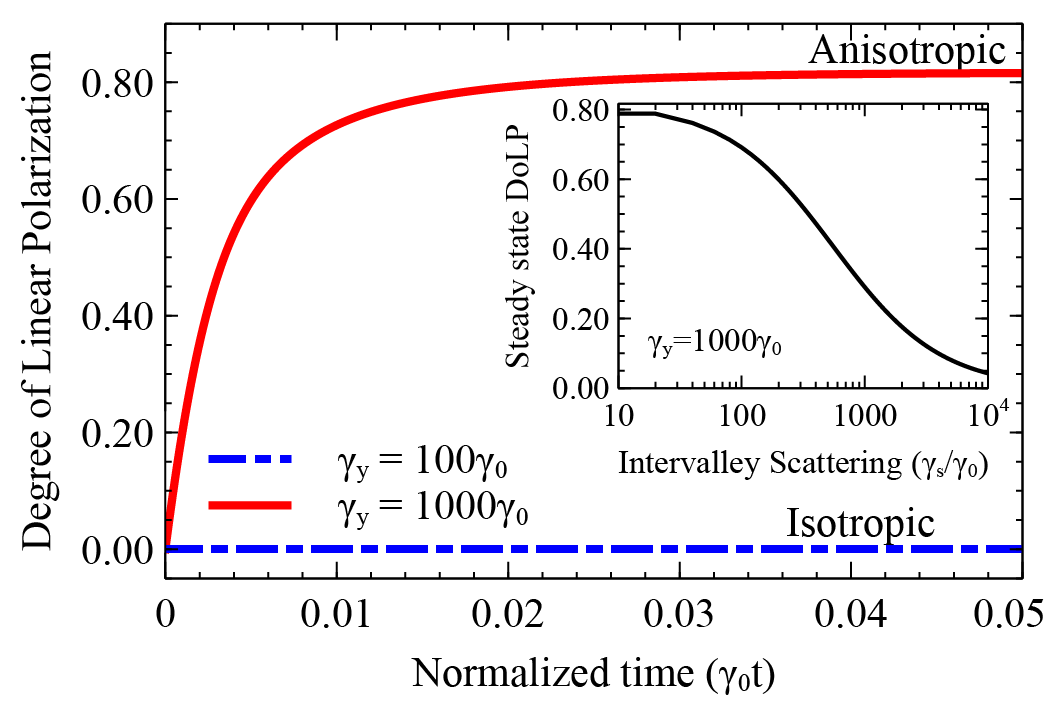}
     \caption{ Temporal evolution of degree of linear polarization (DoLP) for different values of $\gamma_y$ . The inset figure shows the dependence of steady state DoLP with intervalley scattering rate. For the entire figure, $\gamma_x =100\gamma_0$ and pumping rate is taken as  $(R=\gamma_0)$. }
\label{fig:fig2}
\end{figure}
In a heterostructure, the distance between valleytronic material and the anisotropic 2D material can be varied through the use of a spacer layer. This distance has to be optimized for a given frequency and optical conductivity (which is also frequency dependent in general) to efficiently excite the polariton. To obtain higher coherence values, the polariton must be narrowly confined along one direction i.e, $\theta \approx 0\text{ or }90^o$, resulting in an anisotropic electric field enhancement. From Eq.~\ref{eq:disp4} the optimal distances for high coherences are given by $   d = {\sigma''_{xx}}/{(2 \varepsilon_0 \omega)}\text{ or } {\sigma''_{yy}}/{(2 \varepsilon_0 \omega)}$. As shown in Fig.~\ref{fig:fig11}, an overall shift of the valley coherence map is seen as a function of the distance $d$ of the dipole from the metasurface. This is because in order to couple to a surface mode, the in-plane wavevector $q$ must be larger than the gapped region of the isofrequency surface. This means that there is an optimal range of the conductivity tensor for a given distance $d$ where the isofrequency surface of the elliptical or the hyperbolic plasmon intersects\cite{doi:10.1021/acs.nanolett.5b01191}, explaining the distance dependence.
\begin{figure*}[t!]
\includegraphics[width=\linewidth]{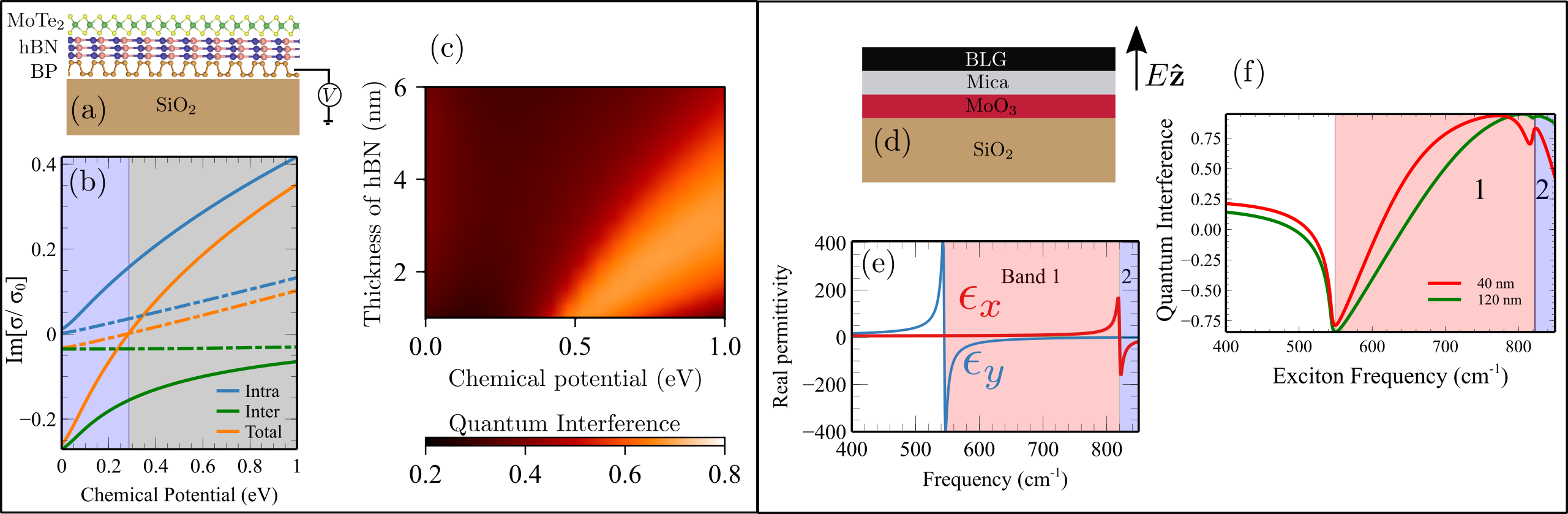}
\caption{(a-c) are for TMDC / hBN / phosphorene heterostructure: (a) Schematic of the heterostructure. (b) Imaginary parts of conductivities of phosphorene (solid and dotted lines are for armchair and zigzag directions respectively) vs. chemical potential($\mu$) at MoTe$_2$ exciton frequency 1.1eV. The blue and grey shaded region represents the elliptical insulator regime and elliptical metallic regime, respectively. (c) Colour plot of spontaneous valley coherence with hBN thickness and chemical potential($\mu$). The Fermi level is defined and $E_f = E_c(\Gamma) + \mu$. The bright region represents the high valley coherence region. (d-f) are for BLG / Mica / $\alpha-$MoO$_3$ heterostructure: (d) Schematic of the heterostructure. (e) Dielectric function of $\alpha-$MoO$_3$. The white region represents elliptical insulating regime. Orange and blue regions represent two hyperbolic bands. (f) Valley coherence as a function of the BLG exciton frequency in the heterostructure.
}
\label{fig:fig3}
\end{figure*}
If the frequency of the exciton is varied, we obtain qualitatively similar trend in the valley coherence map as shown in Fig.~\ref{fig:fig11}(b-d), where we have compared the exciton frequency of 1 eV and 0.1 eV. This is because the minimum conductivity condition is determined only by the product $k_0 d$.  

Experimentally, the spontaneous valley coherence is measured by a metric called the degree of linear polarization (DoLP)\cite{Qiu2019}:
\begin{eqnarray}
   \label{eq:dolp}
   DoLP =\frac{I_{H}-I_{V}}{I_{H}+I_{V}} = \frac{\varrho_{KK^{\prime}}+\varrho_{K^{\prime}K}}{\varrho_{KK} + \varrho_{K^{\prime}K^{\prime}}}
\end{eqnarray}
where $I_H$ and $I_V$ are the intensities of the horizontal and vertical polarized emissions respectively and $\varrho$ is the density matrix. 

The steady state DoLP in the absence of pumping is an ill-defined quantity. Therefore to measure the spontaneous coherence generated, we use a weak incoherent bidirectional pump. The exciton population dynamics are governed by the rate equations (see Supporting Information). Using these rate equations the steady state DoLP can be written as:
\begin{eqnarray}
   \label{eq:ST_dolp}
   DoLP(t\rightarrow \infty) = -\frac{Q}{(1+\frac{R}{\gamma})}
\end{eqnarray}
where $R$ is the bidirectional pumping rate. When the pumping is weak, i.e $R\ll\gamma$, Eq.~\ref{eq:ST_dolp} can be written as $DoLP(t\rightarrow \infty) \approx -Q$.

Fig.~\ref{fig:fig2} shows the temporal evolution of $DoLP$ upon optical pumping. With an isotropic 2D material, the DoLP is zero (shown by the red curve), whereas the presence of an anisotropic 2D material induces a time evolution of the $DoLP$ which approaches a steady state ($\sim -Q$), given by Eq.~\ref{eq:ST_dolp}. Fig.~\ref{fig:fig2} inset shows the dependence of steady state coherence in the presence of intervalley scattering, a decoherence mechanism present between the two valleys. This is attributed to the electron hole exchange interaction and phonon assisted processes\cite{Hao2016,Zeng2012,doi:10.1063/1.4768299,Carvalho2017}.

\emph{Example systems---}
In order to realize the anisotropic quantum vacuum, we propose two layered materials which support anisotropic polaritons. In the following we focus on plasmon and phonon polaritons. In both cases, we propose the tunability of spontaneous valley coherence by active and passive routes respectively.  

As the first example heterostructure for demonstrating spontaneous valley coherence, we consider a naturally in-plane anisotropic material, phosphorene\cite{Xia2019,Carvalho2016} placed near a TMDC which supports valley excitons. Fig.~\ref{fig:fig3}(a) illustrates this system with phosphorene sandwiched between SiO$_2$ substrate and a finite thickness h-BN dielectric\cite{doi:10.1021/acs.nanolett.6b00154}. Monolayer MoTe$_2$\cite{PhysRevB.94.235408,doi:10.1021/nl502557g} is placed on top of the h-BN. The geometrical anisotropy of the unit cell of phosphorene translates to an anisotropic optical conductivity\cite{PhysRevApplied.12.014011,low2014tunable}.
The optical conductivity of phosphorene in the armchair and zigzag directions as a function of chemical potential $\mu$ is shown in Fig.~\ref{fig:fig3}(b). The blue region is the elliptical insulator regime where $\Im\{\sigma_{xx}\},\Im \{\sigma_{xx}\}<0$ and the grey region is the elliptical metallic regime where $\Im\{\sigma_{xx}\},\Im \{\sigma_{xx}\}>0$. This chemical potential modification can be controlled via electrostatic gating\cite{Li2019,Buscema2014,Youngblood2015} or chemical routes\cite{doi:10.1021/acs.nanolett.5b03278,Xiang2015}. Doping provides an easy route to realize an actively tunable natural anisotropic environment for the monolayer TMDC, which enables one to modulate the valley coherence. For very small thicknesses of hBN, one needs to consider nonlocal corrections to the optical conductivity of phosphorene. Recent calculations\cite{PhysRevB.96.205430} have shown however that at our frequency of interest, the spontaneous emission rate is insensitive to nonlocal effects for dipole distances greater than $\sim 2$ nm.

 Fig.~\ref{fig:fig3}(c) illustrates the color plot of the Quantum Interference $Q$ depending on h-BN thickness and chemical potential ($\mu$). We have considered chemical potentials below 1 eV -- a range which has been shown to be achievable with electrostatic doping techniques\cite{doi:10.1021/acsnano.5b00497,Li2019,Buscema2014,Youngblood2015}. For simplicity, we consider the case where the intervalley scattering is negligible $\gamma_s = 0$, which is a reasonable assumption for recently demonstrated high quantum yield functionalized TMDCs \cite{Amani2015,doi:10.1021/acsnano.7b02521}. Fig.~\ref{fig:fig3}(c) further shows that as the thickness of hBN is increased, the interaction of the exciton with the surface plasmon modes decreases, yielding low coherence values. 
The bright region (high $Q$) starts when $\mu \sim 0.3$eV. This can be explained on the basis of the optical conductivity presented in Fig.\ref{fig:fig2}(b). When $\mu<0.3$ eV, the optical conductivity of phosphorene falls under insulating regime yielding low coherence values since in this case only very weakly confined TE modes are supported, see also Fig.~\ref{fig:fig11}(b-d). For $\mu>0.3$ eV, we are in the metallic regime, phosphorene plasmons are excited by the dipole, enhancing the decay rates and the anisotropy, yielding high coherence values as discussed earlier. 

Recently experiments have shown that $\alpha$-MoO$_3$ exhibits in-plane hyperbolicity\cite{Ma2018,Zheng2019} in the mid-infrared range due to anisotropic phonon polaritons. Interestingly bilayer graphene (BLG) excitons can be tuned electrically in the hyperbolic regime of the MoO$_3$\cite{Ju2017}. This is our second proposed heterostructure for realizing tunable valley coherence in $\alpha$-MoO$_3$/BLG heterostructure, as shown in Fig.~\ref{fig:fig3}(d). Here electrical tunability can be achieved by tuning the frequency of the excitons of the BLG via Stark effect. As per our general guidelines in Fig.~\ref{fig:fig11}, high valley coherence in hyperbolic regime occurs in large insulating and low metallic region which is near 800 cm$^{-1}$. Additionally, we also see a large valley coherence at 550 cm$^{-1}$ at the transition region between the elliptical and hyperbolic Band 1 region of MoO$_3$. In Fig.~\ref{fig:fig3}(f) we show the maximum valley coherence vs the exciton frequency for various MoO$_3$ thicknesses. The thickness enters the problem through its dependence on the 2D conductivity of MoO$_3$. As shown in reference \cite{Ma2018}, the in plane polariton wavevector decreases with increasing thickness. The thickness presents another knob for selecting the wavevector at a given frequency and spacer thickness -- thus controlling the spectral location of the valley coherence maxima as shown in Fig.~\ref{fig:fig3}(f). To achieve the exciton energies in BLG as given in Fig.~\ref{fig:fig3}(f), one needs to apply displacement fields ranging from 0.69 to 1.37 V/nm which is well within the experimental reach\cite{Ju2017}. 

\emph{Conclusion---} In summary, we discussed how the optical anisotropy of the polaritons supported by layered material, which is placed near a valleytronic 2D material results in enhanced spontaneous valley coherence in the heterostructure. We explored the phase space of optical conductivity tensor of such an anisotropic 2D material and presented a valley coherence map in the hyperbolic and elliptical regions to guide experimental designs. As example systems, we showed how this valley coherence could be tuned electrostatically in the near infrared using phosphorene/TMDC heterostructure and in the mid-infrared in $\alpha$-MoO$_3$/BLG heterostructure. Our proposed tunable valley coherence in 2D heterostructures offers a natural materials platform for quantum valley physics and applications\cite{PhysRevB.96.245410}.


\emph{Acknowledgement.} A.K. acknowledges funding from Department of Science and Technology grant numbers SB/S2/RJN-110/2017 and DST/NM/NS-2018/49. We thank Kaveh Khaliji for helpful discussions.
\bibliography{references}

\end{document}